# Carbon Nanotube Peapods Under High-Strain Rate Conditions: A Molecular Dynamics Investigation

J. M. De Sousa[1], C. F. Woellner[2], L. D. Machado[3], P. A. S. Autreto[4] and D. S. Galvao[5,6]

[1] *Instituto Federal do Piauí – IFPI, São Raimundo Nonato, Piauí, 64770-000, PI, Brazil.*

[2] *Physics Department, Federal University of Paraná – UFPR, Curitiba, 81531-980, PR, Brazil.*

[3] *Departamento de Física Teórica e Experimental, Universidade Federal do Rio Grande do Norte –UFRN, Natal, 59072-970, RN, Brazil.*

[4] *Center of Natural Human Science (CCNH), Federal University of ABC – UFABC, Santo Andre, 09210-580, SP, Brazil.*

[5] *Applied Physics Department, State University of Campinas – UNICAMP, Campinas, 13083-859, SP, Brazil.*

[6] *Center for Computing in Engineering and Sciences, State University of Campinas – UNICAMP, Campinas, 13083-859, SP, Brazil.*

ABSTRACT

New forms of carbon-based materials have received great attention, and the developed materials have found many applications in nanotechnology. Interesting novel carbon structures include the carbon peapods, which are comprised of fullerenes encapsulated within carbon nanotubes. Peapod-like nanostructures have been successfully synthesized, and have been used in optical modulation devices, transistors, solar cells, and in other devices. However, the mechanical properties of these structures are not completely elucidated. In this work, we investigated, using fully atomistic molecular dynamics simulations, the deformation of carbon peapods under high-strain rate conditions, which are achieved by shooting the peapods at ultrasonic velocities against a rigid substrate. Our results show that carbon peapods experience large deformation at impact, and undergo multiple fracture pathways, depending primarily on the relative orientation between the peapod and the substrate, and the impact velocity. Observed outcomes include fullerene ejection, carbon nanotube fracture, fullerene, and nanotube coalescence, as well as the formation of amorphous carbon structures.

# INTRODUCTION:

Nanostructured systems under high-strain rate conditions have been the subject of theoretical and experimental investigations in recent years. A recent joint theory-experiment study showed that carbon nanotubes (CNTs) and boron nitride nanotubes (BNNTs) under high-velocity impacts can be unzipped into nanoribbons [1]. However, some aspects of the mechanical behaviour of nanostructured systems under high-velocity impacts have not yet been fully investigated. This is especially true for hybrid systems, such as carbon peapods [2]. Carbon peapods consist of C60 fullerenes encapsulated inside SWCNTs (CNT-C60) - see Fig.1 (a). They have been successfully synthesized by various experimental techniques [2,3,4,5], and have interesting chemical and physical properties [6,7,8,9,10]. Following their synthesis, many experimental studies showed the potential of carbon peapods for applications in nanotechnology. These include the use of peapods in transistors [11], in solar cells [12], as hosts for hydrogen storage by the physisorption mechanism [13], in high-performance lithium-ion batteries [14,15], as nano-memory devices [16], as electrode materials for supercapacitors [17], among other applications.

In the present work, we have investigated the deformation behavior of C60 fullerenes encapsulated inside SWCNTs (CNT-C60) under high-strain rate conditions. Our theoretical study was carried out through fully atomistic molecular dynamics (MD) simulations, performed using the Reactive Force Field (ReaxFF) [18] as implemented in the LAMMPS MD code [19]. Our results show that carbon peapods experience large deformation and undergo multiple fracture pathways, depending on their specific impact velocity and their particular orientation relative to the substrate. Observed outcomes include: i) fullerene ejection; ii) nanotube fracture; iii) nanotube fracture with fullerene coalescence; and iv) the formation of amorphous carbon structures, under extreme conditions. These results provide helpful insights in understanding the structural changes and fracture dynamics of hybrid nanostructures under high-strain rate conditions.

# METHODOLOGY:

The high-velocity mechanical impact of carbon peapods was investigated through fully atomistic reactive molecular dynamics simulations using ReaxFF [18]. ReaxFF is a force field that allows for breaking and formation of chemical bonds during the MD simulations. Typically, ReaxFF is parameterized using data from first-principles calculations, and then the simulations use the resulting parameters to generate data that is contrasted against experimental results [18]. It is a modern reactive potential that enables the simulation of large systems with good accuracy at a relatively low computational cost. ReaxFF has been previously used to study the behavior of carbon nanoscrolls at high impact [20,21], as well as the mechanical properties of various nanostructured systems [22].

In our simulations, we used a (12,12) single-wall carbon nanotube, 150 Å long, with a diameter of 16.50 Å, and composed of 2976 carbon atoms. This CNT was filled with ten C60 molecules. We shot the CNT-C60 system at ultrasonic velocities against a rigid substrate, with speed values in the range between 1 km/s and 6 km/s, with a step size of 1 km/s (i. e., we considered six-speed values in total). For each speed value, we considered two relative orientations between the peapods and the target, which we labeled vertical and lateral configurations, see Fig. 1 (b) and (c). For every simulation considered here, we used a time-step of 0.025 fs to integrate the equations of motion. Before shooting a peapod, we equilibrated the system in the NVT ensemble at 10 K for 500000 time-steps, using a chain of three Nosé-Hoover thermostats [23]. A low temperature was used to reduce thermal oscillations, which can add noise to some of the analysis performed here. Still, it is important to remark that tests were also carried out at 300 K, and that impact outcomes remain almost the same. After the end of the equilibration, we turned the thermostat off and shot the peapods at supersonic velocities against a solid target, and then let the system evolve freely in the NVE ensemble for another 500000 time-steps.

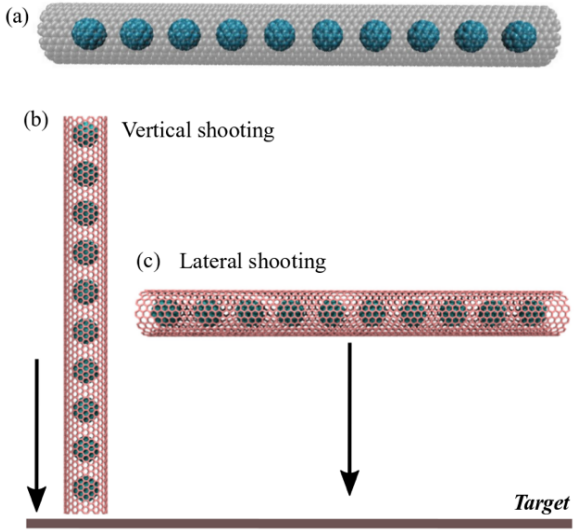

**Figure 1:** Atomic model and impact orientations of carbon peapods. (a) carbon peapods, (b) vertical axis orientation and (c) lateral axis orientation.

**RESULTS AND DISCUSSION:**

In Fig. 2 we present the temporal evolution (for the lateral impacts) of the percentage of covalent bond formation between CNT and $C_{60}$. Up to 4 km/s only a few new covalent bonds are formed. After 4 km/s a significant number of new covalent bonds are formed with a tendency to saturate. For the case of 6 km/s the saturation regime has not yet been reached. It is interesting to notice the presence of some peaks, especially for the 4 km/s case. This can be explained as an elastic reconstructions with some covalent bonds being broken and the $C_{60}$ recovering its initial structural configurations. As the kinetic energy is increased these processes become more difficult to occur.

The high kinetic energy (especially for 4 km/s up to 6 km/s) causes several structural failures (fractures) both in the carbon nanotubes and fullerenes, leading some time to the formation of amorphous structures (see discussions below). At a speed of 6 km/s we can observe a percentage of bond formation between CNT and $C_{60}$ as high as approximately 16%.

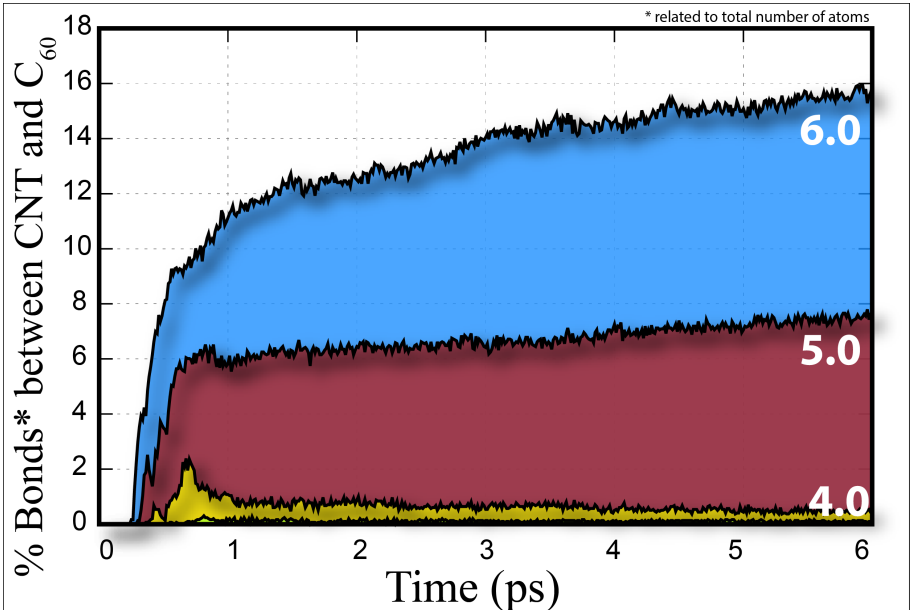

**Figure 2:** Time evolution of the percentage of the covalent bonds formed between CNT(12,12) and C60 as a function ultrasonic velocities values.

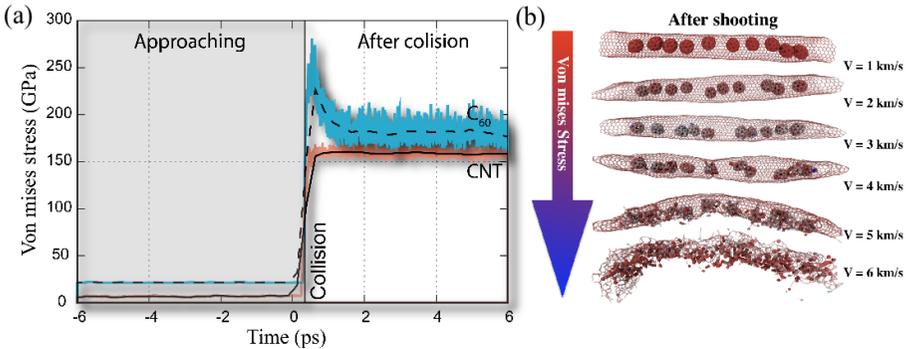

**Figure 3:** (a) von Mises stress (GPa) as a function of simulation time for a peapod impact speed of 5km/s. (b) Representative MD snapshots depicting the resulting structures after impacts as a function of velocity values. The color in the snapshots indicate the local stress values with red (darker) and blue (lighter) representing low and high stress regimes, respectively. The colors of the arrow provides a more detailed scale, whereas its head indicates the substrate direction.

In Fig. 3(a), we present the von Mises stress values as a function of temporal evolution (before and after impact). From this Fig. we can notice that the C60 are under a higher stress level than the nanotube. In Fig.3(b) we present representative MD snapshots after impacts at different velocities. The snapshots show the peapod structural transitions, from its initial configurations (1 km/s), followed by C60 rearrangements (up to 3 Km/s), structural failures (fractures) (up to 5 Km/s) and then complete amorphization (6 Km/s).

The presence of C60 significantly affects the CNT deformation mechanisms, as illustrated in Fig. 4. For the empty CNT, the impact induces a radial tube collapse [1], which results in the CNT unzipping. However, the presence of C60 in the peapod blocks such deformations and the tube remains almost intact and under a lower stress level.

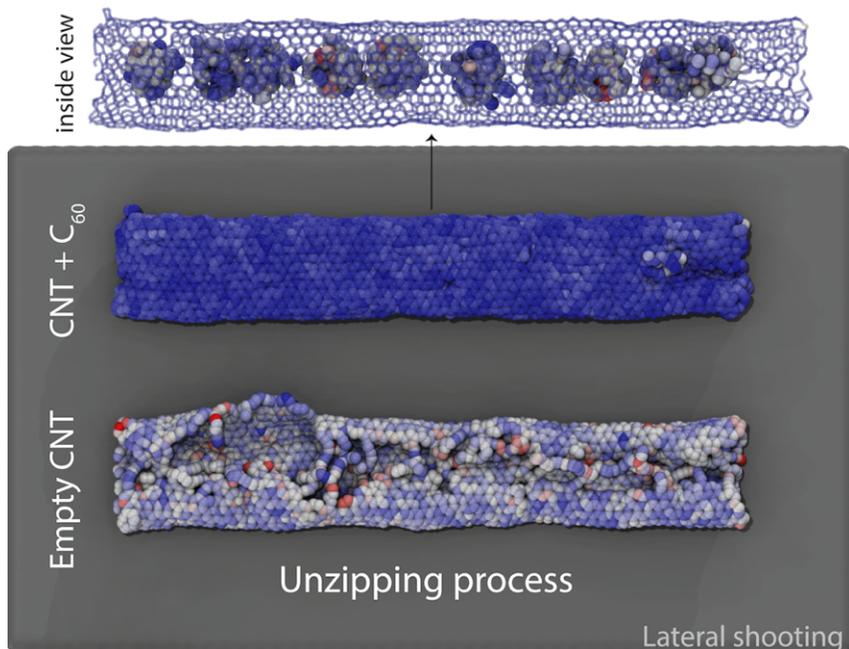

**Figure 4:** Representative MD snapshots of deformation behavior of peapods and empty carbon nanotube for the case of 5 Km/s. The color indicates the local stress values, where blue (lighter) and red (darker) indicate low and high-stress regimes, respectively.

In Fig. 5 we present representative MD snapshots for vertical shootings at 4 Km/s. In this case we observe typically C60 ejection and/or carbon amorphization. Structural amorphization for vertical shootings were also reported for CNT [1], carbon and boron nitride nanoscrolls [21] and boron nitride nanotubes [25].

## CONCLUSIONS:

We have investigated through fully atomistic reactive molecular dynamics (MD) simulations the dynamics of high velocity (ultrasonic) ballistic impacts of peapods (CNT with encapsulated C60 molecules) against solid targets. We considered the cases of vertical and horizontal shootings. Our MD results show that peapods exhibit remarkable resilience under high-strain regimes. Our results show that carbon peapods can experience large structural deformations and undergo multiple fracture pathways, depending on their specific impact velocity and their particular orientation relative to the substrate. Observed outcomes for the lateral impacts include: i) fullerene ejection; ii) nanotube fracture; iii) nanotube fracture with fullerene coalescence; and iv) the formation of amorphous carbon structures. For the vertical impacts, we observed mainly C60 ejections and carbon amorphizations. The presence of C60 inside the tube significantly affects its deformation mechanisms blocking the tube radial collapse and allowing the tube to experience larger strain levels before fracturing. These results provide helpful insights in understanding the structural changes and fracture dynamics of hybrid nanostructures under high-strain conditions.

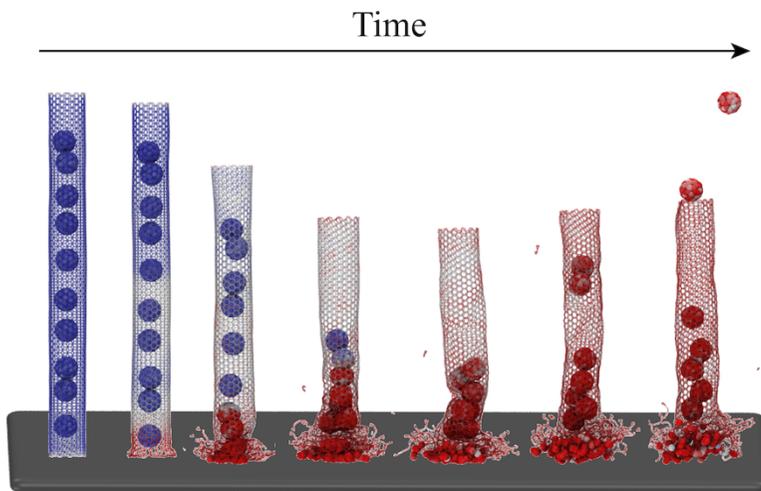

**Figure 5:** Representative MD snapshots of the peapod deformation behavior for vertical shootings and velocity of 4 Km/s. The color indicates the local stress values, where blue (lighter) and red (darker) indicate low and high-stress regimes, respectively.

**ACKNOWLEDGMENTS:**

This work was supported in part by the Brazilian Agencies CAPES, CNPq, FAPESP. J.M.S, C.F.W, L.D.M., P.A.S.A and D.S.G thank the Center for Computational Engineering and Sciences at Unicamp for financial support through the FAPESP/CEPID Grant #2013/08293-7.